\documentclass[conference]{IEEEtran}
\pdfoutput=1 
\usepackage{amsmath}
\usepackage{csquotes}
\usepackage{caption}
\usepackage{subcaption}
\usepackage{url}
\usepackage{breakcites}
\usepackage{multirow}
\usepackage{makecell}
\usepackage{ragged2e}
\usepackage{multirow}
\usepackage{makecell}

\usepackage[figuresright]{rotating}
\usepackage{setspace}
\usepackage{booktabs}
\usepackage{enumerate}
\usepackage{pdflscape}\usepackage{rotating}

\usepackage{verbatim}

\usepackage{hyperref}

\title{A Cradle-to-Gate Life Cycle Analysis of Bitcoin Mining Equipment Using Sphera LCA and ecoinvent Databases}
\author{\IEEEauthorblockN{Ludmila Courtillat-{}-Piazza$^{a,b,*}$, Thibault Pirson$^b$, Louis Golard$^b$, David Bol$^{b}$}
\IEEEauthorblockA{
\textit{$^a$ENS Rennes}, Université de Rennes, France, \\
\textit{$^b$ICTEAM}, Université catholique de Louvain, Belgium} \\
$^*$\textit{Corresponding author at:} Bâtiment Maxwell a.193, Place du Levant, 3 (L5.03.02) 1348 Louvain-la-Neuve, Belgium.\\
\textit{Email address: ludmila.courtillat-piazza@ens-rennes.fr, \{thibault.pirson,louis.golard,david.bol\}@uclouvain.be}\\
}

\begin{document}
\maketitle


\justifying
\textbf{Abstract} Bitcoin mining is regularly pointed out for its massive energy consumption and associated greenhouse gas emissions, hence contributing significantly to climate change.
However, most studies ignore the environmental impacts of producing mining equipment, which is problematic given the short lifespan of such highly specific hardware.
In this study, we perform a cradle-to-gate life cycle assessment (LCA) of dedicated Bitcoin mining equipment, considering their specific architecture.
Our results show that the application-specific integrated circuit designed for Bitcoin mining is the main contributor to production-related impacts.
This observation applies to most impact categories, including the global warming potential.
In addition, this finding stresses out the necessity to carefully consider the specificity of the hardware.
By comparing these results with several usage scenarios, we also demonstrate that the impacts of producing this type of equipment can be significant (up to 80\% of the total life cycle impacts), depending on the sources of electricity supply for the use phase.
Therefore, we highlight the need to consider the production phase when assessing the environmental impacts of Bitcoin mining hardware.
To test the validity of our results, we use the Sphera LCA and ecoinvent databases for the background modeling of our system.
Surprisingly, it leads to results with variations of up to 4 orders of magnitude for toxicity-related indicators, despite using the same foreground modeling.
This database mismatch phenomenon, already identified in previous studies, calls for better understanding, consideration and discussion of environmental impacts in the field of electronics, going well beyond climate change indicators.\\


\textbf{Warning:} This document is a pre-print version of the article: ``A Cradle-to-Gate Life Cycle Analysis of Bitcoin Mining Equipment Using Sphera LCA and ecoinvent Databases''.
\section{Introduction}

For more than a century, human activities have exerted ever-increasing pressures on climate and ecosystems \cite{masson2021ipcc}\cite{steffen2015trajectory}. Such pressures should be limited to avoid dramatic and irreversible consequences on Earth and hence on Humanity \cite{steffen2015planetary}. Therefore, all economic sectors must urgently reduce their environmental impacts, and in particular their greenhouse gas (GHG) emissions \cite{masson2021ipcc}\cite{agreement2015paris}.
In this context, cryptocurrency mining, and especially Bitcoin mining, has been several times flagged up for its emissions of several tens of MtCO$_2$eq per year \cite{krause2018quantification}, which are still rising so far \cite{index2023ghg}.
These emissions have been compared to the annual GHG emissions of countries like Belgium \cite{de2020bitcoin}, Poland \cite{index2020comparison}, or Sri Lanka \cite{stoll2019carbon}. 
They are mainly due to the \emph{proof-of-work} consensus mechanism, which is a computation-intensive process requiring important energy consumption \cite{de2018bitcoin}.
This mechanism is used in the blockchain of several cryptocurrencies, including Bitcoin which is the most popular cryptocurrency. Hence, it generates much more activity than other proof-of-work-based cryptocurrencies such as Litecoin or Dogecoin.
For instance, the authors of \cite{gallersdorfer2020energy} estimated that Bitcoin was responsible for 68\% of the total energy consumption of the top-20 mineable currencies in July 2020.

In addition to electricity, cryptocurrency mining also requires highly specialized and efficient hardware to implement the consensus mechanism.
To be economically profitable, a cryptocurrency mining equipment must cost less money in energy consumption than it earns by mining the cryptocurrency.
The profitability of such equipment is determined by comparing its efficiency to the global computational capacity of the cryptocurrency network, where the efficiency is the computational rate of the mining equipment (the hashrate) divided by its power consumption \cite{taylor2013bitcoin}.
Indeed, once the global average efficiency of the cryptocurrency network increases, the profitability of the miner decreases.
Consequently, each generation of mining equipment quickly becomes obsolete (after 1.5 year on average according to \cite{de2021bitcoin}), leading to frequent replacement by the owner.
Cryptocurrency mining hardware must therefore be highly specialized to implement efficiently the consensus mechanism, and highly efficient to be profitable.
For a decade, almost all Bitcoin mining operations have been carried out using specialized equipment based on application-specific integrated circuits (ASICs), hereafter referred to as \textit{ASIC miners} \cite{taylor2013bitcoin}\cite{taylor2017evolution}.

Because of the high environmental impact of electronic device production \cite{clement2020sources}\cite{pirson2022environmental}, the fast renewal cycle of the ASIC miners raises concerns. Hence, this study aims at evaluating the impacts of producing an ASIC miner.

\subsection{Related works}

Previous scientific studies have investigated the use-phase energy consumption and related GHG emissions of Bitcoin mining during the computation process.
A first paper addressing this concern was published in 2014 \cite{odwyer2014bitcoin}, followed by other ones, increasingly detailed on the modeling of miners' geographical distribution, miners' specifications and market dynamics.
For instance, since 2018, a strong effort was deployed by De Vries, Stoll and colleagues to address this question \cite{de2018bitcoin}\cite{stoll2019carbon} \cite{de2020bitcoin} \cite{gallersdorfer2020energy} \cite{de2022revisiting}.
Other studies with similar approaches were also conduced by other authors \cite{krause2018quantification}, \cite{li2019energy}, including work in the grey literature \cite{bevand2017electricity}.
While many studies focus on the energy consumption of Bitcoin mining, there are in contrast few investigations addressing the embodied environmental impacts of the hardware used for mining. 
To the best of our knowledge, only one study estimates the global mass of e-waste generated by Bitcoin mining activities between 2014 and 2021 \cite{de2021bitcoin}.

Moreover, the life cycle assessment (LCA) methodology for assessing environmental impacts has been scarcely used for studying Bitcoin mining, despite the extensive literature on its climate impact. Indeed, a majority of the studies adopt a market approach, by combining material specifications and macro indicators to assess the Bitcoin network impacts as in \cite{de2018bitcoin}, \cite{bevand2017electricity} or \cite{de2020bitcoin}. In addition, these studies mainly focus on the energy consumption and related GHG emissions of Bitcoin mining and do not include a wider range of environmental impact categories and life cycle phases, in contrast to a life cycle approach.
A streamlined literature review
revealed only five studies mentioning \textit{LCA} and \textit{Bitcoin}, among which only two include the production phase in the LCA.

More precisely, \cite{parrercomparative}, \cite{pagone2023carbon} and \cite{pizzol2021non} deal with an LCA of a Bitcoin mining activity, but either they do not carry out the LCA themselves, or they exclude the production phase from their scope.
In \cite{kohler2019life}, the authors 
perform a cradle-to-grave LCA on the whole Bitcoin network, excluding the transport phase.
Similarly, \cite{roeck2022life} conduct a cradle-to-grave LCA of Bitcoin mining, but at the scale of a power plant in the United States, without considering transport nor end-of-life.
Both studies obtain similar conclusions and claim that the production phase represents less than 1\% of GHG emissions over the whole life cycle of the system they analyzed.
Yet, as electronic hardware production is not the focus of these studies, it was modeled in both cases by selecting a generic process in an LCA database representing the production of a typical computer and scaled based on the mass of the ASIC miner.
Hence, we argue that these studies significantly overlook the specificity of mining equipment.

In summary, we highlight some gaps and weaknesses in the current literature about the environmental impacts of crypto-mining: (i) the small number of LCA studies dealing with Bitcoin and crypto-mining, (ii) the absence of consideration for the specificity of crypto-mining equipment, (iii) a lack of analysis of embodied impacts, except for the climate change impact category.

Therefore, this paper aims at filling this gaps.

\subsection{Research contributions}

In the previous section, we highlight that it is necessary to model more precisely Bitcoin mining equipment as it has a specific architecture comprising numerous ASICs. It also includes large active and passive cooling systems to prevent overheating.
In addition, we point out the need of a multi-indicators approach, as focusing on a subset of environmental indicators can hide burden-shifting issues.
Consequently, this study aims at assessing the environmental impacts of ASIC miners’ production using a multi-indicators perspective, while taking into account their specific architecture.

More specifically, we present three main contributions: 
\begin{enumerate}[(i)]
    \item an in-depth multi-indicators cradle-to-gate attributional LCA of the Antminer S9 Bitcoin ASIC miner, with the APW3++ power supply. 
    \item A comparison of the production and the use phases of the mining equipment, taking into account various scenarios of lifespan and geographic locations. 
    \item A comparison of the obtained results using two different LCA databases, namely Sphera LCA and ecoinvent.
\end{enumerate}

To the best our knowledge, this work proposes the first LCA taking into account the specificity of ASIC miners. We expect contribution (i) and (ii) to be of interest for the community investigating the environmental impacts of Bitcoin, and contribution (i) and (iii) for the LCA community in general.

\subsection{Structure}

The rest of this paper is structured as follows.
Section \ref{sec:methods} presents the goal, the scope and the modeling principles of the LCA study, Section \ref{sec:results} provides and interprets the results of the LCA study and discusses the differences between the two databases. Then, Section \ref{sec:limiations-and-uncertainty} discusses the limitations and uncertainty of the LCA study.
Section \ref{sec:beyond-carbon} puts the results in perspective by comparing the production and the use phases for different scenarios in a multi-indicators perspective before the conclusion in Section \ref{sec:conclusion}.

\section{Method}
\label{sec:methods}

\begin{figure*}
    \centering
    \includegraphics[width=0.7\textwidth]{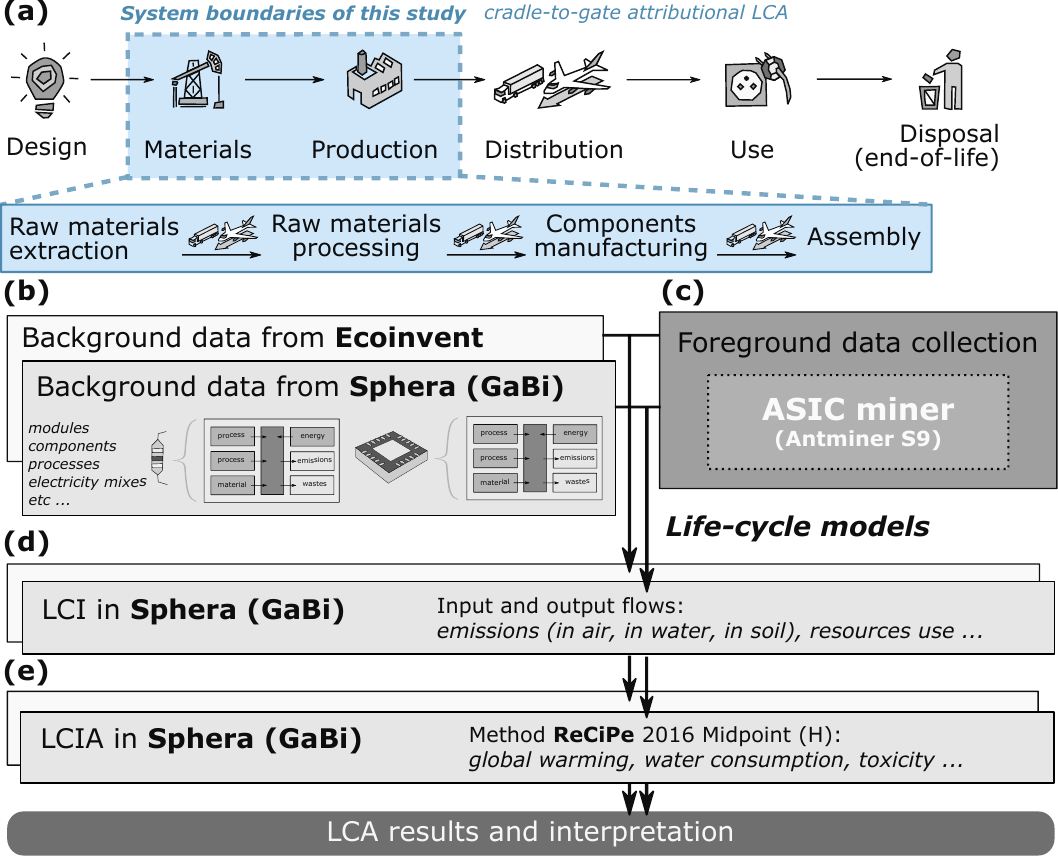}
    \vspace{0.1cm}
    \caption{Methodology of the LCA study, including (a) the system boundaries, (b) the two parallel background data collection using Sphera LCA (GaBi) and ecoinvent databases, (c) the foreground data collection for the ASIC miner modeling (presented in Figure \ref{fig:product-data-model} and detailed in Support Information SI 2), (d) the computation of the LCI flows, and (e) the computation of the LCIA results using ReCIPe 2016.}
    \label{fig:methodo}
\end{figure*}

This section describes the goal, the scope and the modeling principles of this study, following the LCA methodology as proposed by \cite{hauschild2018life} in its chapters 7 and 8.
Figure \ref{fig:methodo} summarizes the methodology.

\begin{figure*}
    \centering
    \includegraphics[width=0.7\textwidth]{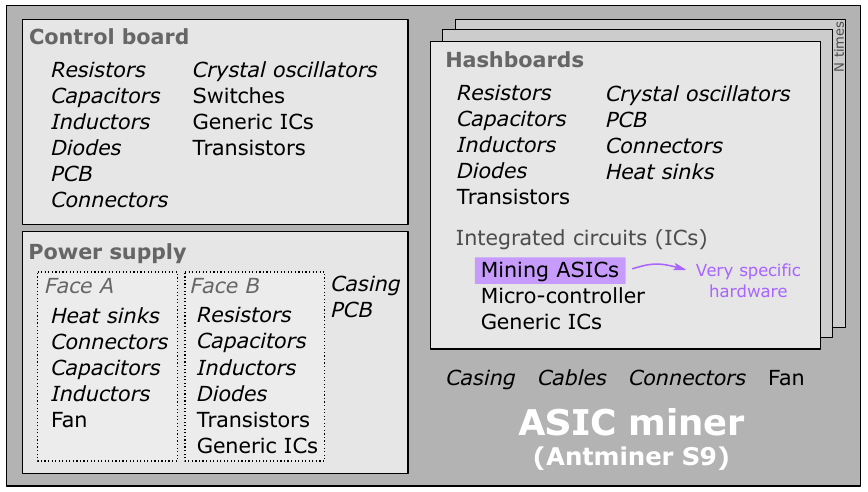}
    \vspace{0.1cm}
    \caption{Architecture of the ASIC miner including a control board, several hashboards and a power supply. In the Antminer S9, there are 3 hashboards with 63 ASICs each, and twice as many heat sinks. Hashboards are used to parallelize the proof-of-work protocol calculations, and the control board distributes the calculations between hashboards and communicates with the network.}
    \label{fig:product-data-model}
\end{figure*}

\subsection{Goal and scope}
\label{subsec:goal-and-scope}

\textbf{\emph{Goal.}} The intended application of our results is to identify the hotspots of environmental impacts in the production of Bitcoin mining equipment. This is a descriptive study which is not directly expected to support decisions causing minor or structural changes.
Its target audience is composed of scientists interested in environmental impacts of information and communication technologies (ICT), more precisely: LCA practitioners and researchers studying the environmental impacts of cryptocurrencies.

\textbf{\emph{Scope.}}
The object under study in this LCA is ASIC miners. We focus on the Antminer S9 which has been the most common Bitcoin miner during several years \cite{coinshare2022the} and that is also known as the mining equipment with the longest lifespan \cite{de2021bitcoin}. In 2018, the Antminer S9 represented 78\% of the active equipment in the Bitcoin network \cite{bendiksen2018bitcoin}.
The functional unit of the study is \emph{the computing activity over the whole operating lifespan of a Bitmain Antminer S9 supplied with an APW3++ power supply.}
We restrict the boundaries of our study to a cradle-to-gate perspective, and we use an attributional modeling framework.
In order to analyze the influence of LCA databases on the results, we use independently the Sphera LCA (former GaBi) and ecoinvent v3.8 databases for the background modeling of the life cycle inventory (LCI), while keeping the same foreground modeling.
The foreground model is transparently provided in the supporting information SI 2, and we also provide the reference of each unit process used for the background model in supporting information SI 1.
The results of the life cycle impact assessment (LCIA) are obtained using the multi-indicators ReCiPe 2016 v1.1 Midpoint (H) method.
Some cut-offs are also applied. First, in both databases, the final assembly and packaging of the miner are excluded. In addition, the absence of some components in databases leads to the exclusion of the connectors when using Sphera LCA, and of crystal oscillators when using ecoinvent v3.8.
In the Sphera LCA case, we perform a sensitivity analysis by implementing three different scenarios covering a low, typical and high scenario for some key parameters.

\subsection{Modeling principles}

Throughout this study, several assumptions were made about the background and foreground modeling of the ASIC miner and its power supply.
The foreground system includes all the processes of the physical value chain. It mainly focuses on the inventory of components present in the hardware, described using their nature, quantity, size, etc. The background modeling consists in modeling the processes of the foreground system with items from the databases, used as ``proxies'' for the components and processes identified in foreground modeling. This step was performed twice using Sphera LCA and ecoinvent v3.8 (market for, APOS | S).
This process is summarized in Fig. \ref{fig:methodo} and further details are available in the supporting information SI 1 and SI 2.

The architecture of an ASIC miner is presented in Fig. \ref{fig:product-data-model} (a control board, three hashboards, a fan, a casing, cables and an independent power supply with its own casing, fans and cables).
This is a simplified representation of the product data model used in this study, which is highly detailed in the support information SI 1 (with pictures) and SI 2.

This model is built using the following guidelines.
When necessary, size, weight, number of pieces and magnetic properties are measured for all components in the control board and for all components in the APW3++ power supply and for the heat sinks used in the hashboards.
In particular, all bulky components (electrolytic capacitors, ring core coils, etc) in the power supply are disassembled for weighting and measurement. Smaller components, such as surface mounted devices (SMD) capacitors or resistors, are only visually inspected and measured.
For the remaining parts (hashboard without ASICs and without heat sinks, casing, cables), the foreground model was established from pictures, videos and information provided by constructors and vendors on the Internet (all the sources are available in SI 2).
Table \ref{tab:modeling-principales} presents an overview of the methods used for data collection of each components and processes. The modeling principles applied to build the background models in both Sphera LCA and ecoinvent v3.8 databases are also detailed in the table.

    \begin{sidewaystable*}
        \caption{Foreground and background modeling principles and assumptions for the LCI models using Sphera LCA and ecoinvent databases. It covers the production of each components of the ASIC miner.
            See supplementary material SI 1 (background) and SI 2 (foreground) for more details.}
        \label{tab:modeling-principales}
        \large
        \centering
        \begin{singlespace}
            \resizebox{1\textwidth}{!}{
                \begin{tabular}{|m{0.05\textwidth}|m{0.5\textwidth}|m{0.45\textwidth}|m{0.45\textwidth}|}
                    \hline
                                             & \multicolumn{1}{c|}{Foreground Data}                                                                                                                                                            & \multicolumn{2}{c|}{Background Data}                                    \\
                    \cline{3-4}
                                             &                                                                                                                                                                                                 & \multicolumn{1}{c|}{Sphera LCA (GaBi)} & \multicolumn{1}{c|}{Ecoinvent} \\
                    \hline
                    \rotatebox{90}{ICs}      &
                    \multirow{2}{*}{\parbox{0.5\textwidth}{\vspace{-0.5cm} The nature of these components (IC, resistor, capacitor ...) is visually identified using PCB labels.
                            Their type is visually identified, supported by the datasheet when available.
                            The die area of the mining ASICs (BM1387B) and of the largest ICs of the control board is measured with a microscope after chemical removal of the packages.
                            The amount of each component is estimated using pictures as well as a teardown, as detailed in the supplementary material.}}
                                             &
                    Unit processes are selected based on this order of importance: 1) the IC function (logic, Flash or DRAM memory), 2) the technological node (when it is available), 3) the package type (BGA, SO, QFN ...), 4) the package size and the number of pins.
                    The unit process is scaled based on the IC die size if this is known, otherwise it is scaled based on the package size.
                                             &
                    Only two types of IC are available: logic or memory.
                    The scaling method relies on the IC mass which is determined by dividing the die area by the wafer area used to produce 1 kg of ICs.
                    This approach has been used by \cite{pirson2022environmental}\cite{clement2020sources}.
                    When the die area is unknown, it is estimated using the die-to-package ratio provided by GaBi documentation for the respective package type.
                    \\
                    \cline{1-1} \cline{3-4}
                    \rotatebox{90}{\parbox{2.1cm}{\centering Passive                                                                                                                                                                                                                                                     \\ components}} &
                                             &
                    Unit processes are selected first according to the type of components, and then to their size. For large components (mainly in the power supply), the scaling is based on their mass.
                                             &
                    Unit processes are selected according to their nature and type. Small components are scaled based on the mass of corresponding processes in the GaBi model.
                    Larger components are scaled based on their real mass.
                    \\
                    \hline
                    \rotatebox{90}{Fan}      &
                    The fan of the Antminer S9 is modeled by disassembling similar fans, especially that of the APW3++.
                                             & Plastic parts of the fan are modeled using \textit{"Polyethylene linear low density granulate"} scaled based on their mass.
                    To model the rest of the fan, we reuse the model of a simplified DC motor from (\cite{pirson2021assessing}).
                                             & We use the specific unit process \textit{"Fan production, for power supply unit, desktop computer"}, scaled based on the mass of the fan.
                    \\
                    \hline
                    \rotatebox{90}{PCB}      &
                    The control board PCB is made of 6 layers, which was observed using a microscope.
                    The hashboards and power supply PCBs are assumed to be made of 2 layers only as their component density does not justify a higher number of layers.
                                             &
                    \multicolumn{2}{m{0.93\textwidth}|}{We select unit processes corresponding to FR4 PCB assembled with the hot air solder leveling method with the desired number of layers.
                        For 6-layers PCBs, we combine 50 \% of 4-layers PCBs with 50 \% of 8-layers PCBs.
                    The scaling is based on the PCB areas.}                                                                                                                                                                                                                                                              \\
                    \hline
                    \rotatebox{90}{\centering\parbox{2cm}{\centering Casing and                                                                                                                                                                                                                                          \\ Heat sinks}} &
                    Metal types are assessed by testing their magnetic and oxidation properties and calculating their mass density.
                    The casing of the Antminer S9 and the power supply is made of steel. Their masses are estimated by subtracting the masses of other modules from their total mass.
                    The heat sinks are made of aluminium.
                                             &
                    The casing and heat sinks are respectively modeled using \textit{"Section bar rolling, steel"} and \textit{"Aluminium extrusion profile"}.
                    The scaling is based on their mass.
                                             &
                    The casing and heat sinks are respectively modeled using \textit{"Steel sheet stamping and bending (5\% loss)"} and \textit{"Aluminium extrusion profile"}.
                    The scaling is based on their mass.
                    \\
                    \hline
                    \rotatebox{90}{Assembly} &
                    SMD and THT components are observed on all board without any further information available to model the assembly processes.
                                             & \multicolumn{2}{m{0.93\textwidth}|}{Solder paste based on Sn, Ag, and Cu (SnAg3....) is selected and both SMD and THT assembly line of the board.
                    }
                    \\
                    \hline
                    \rotatebox{90}{Cables}   &
                    Cables are modeled using online information, except for the PCI-e power cable of the power supply.
                                             & \multicolumn{2}{m{0.93\textwidth}|}{Unit processes are selected based on whether they are power or data cables, and according to their thickness, their number of wires and the mass per meter.
                        The scaling is based on their length.}
                    \\
                    \hline
                    \rotatebox{90}{Others}   & No information are available on upstream transport.
                    Connectors were studied similarly as passive components.
                                             & The upstream transport is included by reproducing the transport assumptions of "market for" processes in ecoinvent (20km of train and 20km of truck).
                    The connectors are not modeled.
                                             & The upstream transport is modeled using  "market for" processes. Connectors are partially modeled.
                    \\
                    \hline
                \end{tabular}
            }
        \end{singlespace}
    \end{sidewaystable*}

\section{Results}\label{sec:results}

This section presents the LCIA results and discusses the differences obtained with the two databases.

\subsection{Hotspots analysis}

\begin{figure*}
    \centering
    \includegraphics[width=\textwidth]{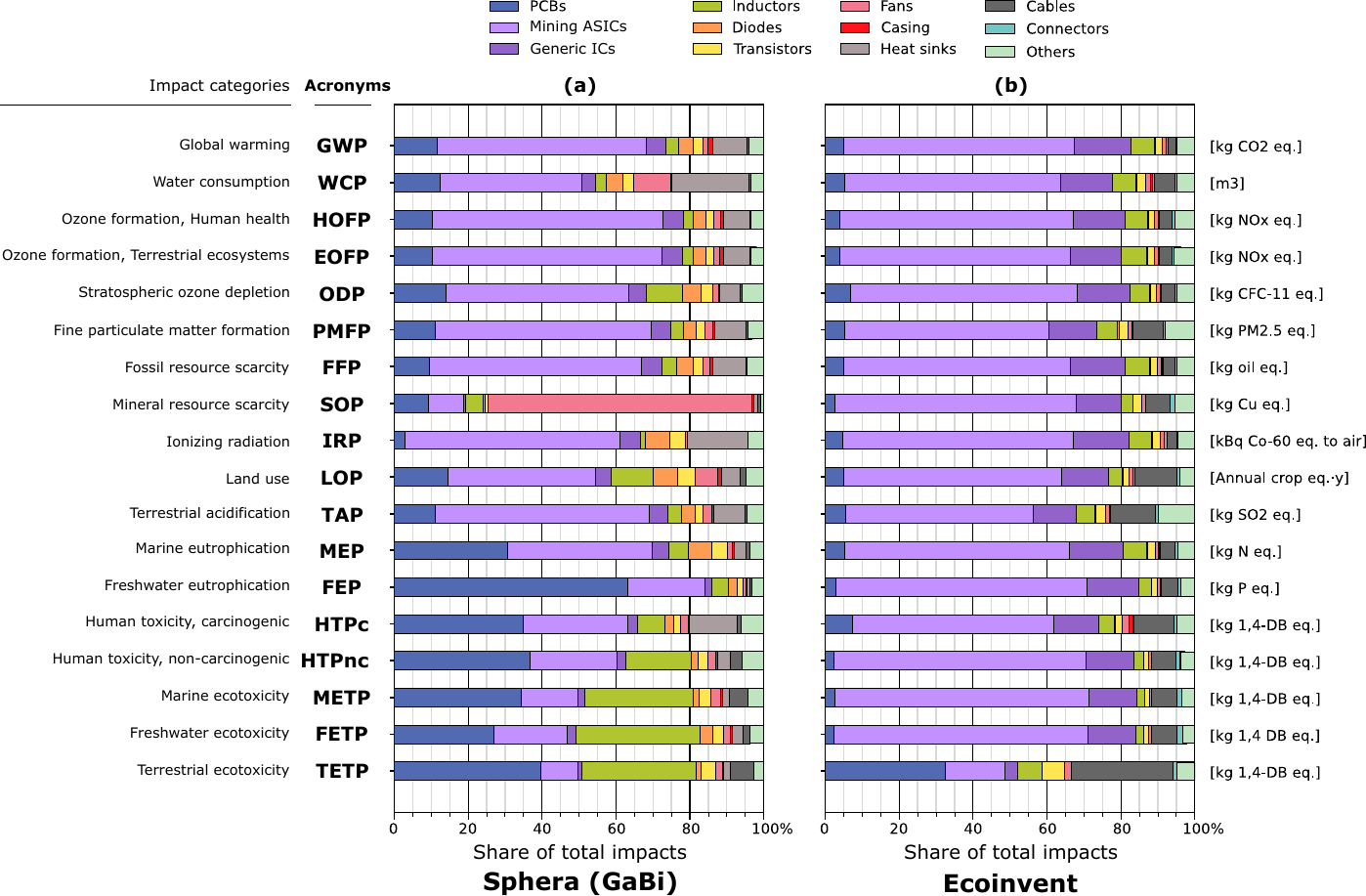}
    \caption{Comparison of the LCIA results (cradle-to-gate) on Antminer S9 miner with APW 3++ power supply, using Sphera LCA (GaBi) and ecoinvent databases. ReCiPe 2016 (H) is the LCIA method used in this study. Data available in the Sheet ``data\_from\_figure\_3\_in\_manuscript'' of the Supporting Information SI 3.}
    \label{fig:results}
\end{figure*}

Figure \ref{fig:results} compares the results obtained using the two databases for each impact category of the LCIA method.
We clearly observe that the production of ASICs is the hotspot in both cases. This is not surprising, as there are 189 ASICs in the whole miner and as ICs are known to be resource- and energy-intensive components, contributing significantly to the impacts of electronic devices production \cite{clement2020sources}\cite{Fair2020life}\cite{pirson2022environmental}.

The printed circuit boards (PCBs) also appear to be an important contributor for terrestrial ecotoxicity in both cases, and for all human and eco-toxicity categories as well as water eutrophication when using Sphera LCA.
Fans have an important impact (roughly 70 \%) on mineral scarcity with the Sphera LCA database, while this is not the case on the ecoinvent side. This difference might be explained by the presence of neodymium in the \emph{custom} model of fan motor used in Sphera LCA, which is not included in the model of a power supply fan in ecoinvent.
Neodymium is a rare earth metal that is used to manufacture powerful magnets, such as those present in fan motors. However, we have no data to confirm whether or not neodymium is present in the fans of the ASIC miner.
The impacts of inductors, cables and diodes can also be noticed, especially regarding toxicity and ecotoxicity impact categories. These impacts can be partly explained by the fact that inductors and cables are mainly made of copper. Indeed, by examining the elementary flows, we observed that emissions of copper to air are responsible for 81 \% of terrestrial ecotoxicity in the Sphera LCA case, and for 73 \% in the ecoinvent one.
PCBs, which are a major contributor to toxicity and ecotoxicity indicators, also contain copper.
Finally, heat sinks, made of aluminium, have a significant impact on several indicators (e.g., water consumption, ionizing radiation, etc) in the Sphera LCA case, but not in the ecoinvent one.

\subsection{Differences between databases}
\label{sec:databases-inconsistencies}

Apart from identifying ASICs as an hotspot, the two databases reveal different shares of total impacts between the components, in spite of a common foreground data collection and product data model. Let us point out and explain these differences.

First, we notice that, in the ecoinvent case, the distribution of impacts is almost always the same for each impact category, except for terrestrial ecotoxicity.
The impacts distribution in Sphera LCA results is much more diverse across the different impact categories.
This may suggest that, in the ecoinvent case, a small number of processes involved in the production of all components of the Antminer S9 are responsible for the majority of impacts in almost all categories.
Therefore, the impacts of each component would be proportional to the quantity of such dominant processes implied in their production.
For instance, these processes may be the raw material extraction or the electricity production, as electricity is required at various manufacturing stages.

\begin{figure*}
    \centering
    \includegraphics[width=0.95\textwidth]{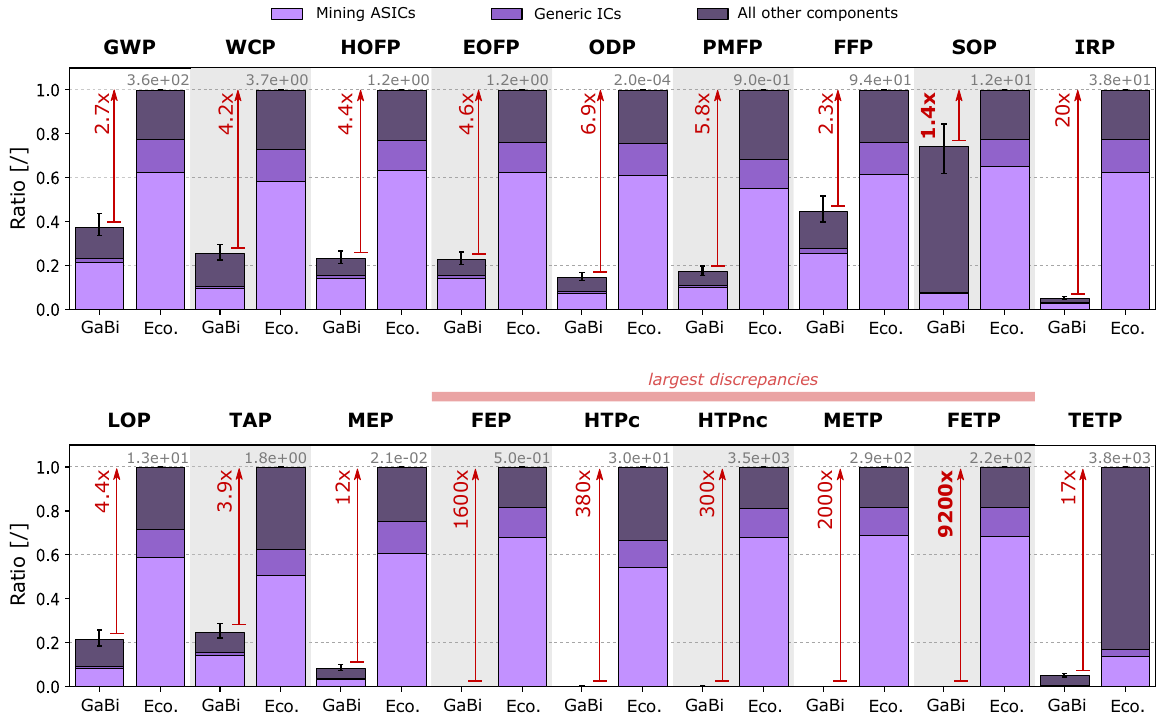}
    \caption{Discrepancies between the absolute LCIA results obtained with Sphera LCA (GaBi) and ecoinvent. Results are normalized with respect to the ecoinvent ones, i.e., ecoinvent = 1 in each impact category. The ecoinvent results are labeled with their real value obtained with the standard units for each impact category. The abbrevations and units are presented in Fig. \ref{fig:results}. The errorbars displayed in the Sphera LCA results are the results of the sensitivity analysis. Data available in the Sheet ``data\_from\_figure\_4\_in\_manuscript'' of the Supporting Information SI 3.
    }
    \label{fig:database-comparison}
\end{figure*}

Second, Figure \ref{fig:database-comparison} compares the absolute LCIA results obtained with Sphera and ecoinvent databases.
For all impact categories, the results obtained with ecoinvent significantly exceed the results with Sphera LCA. In most cases, this can primarily be explained by the fact that the production of ASICs and other ICs is the hotspot of our LCA, and that the impacts of IC production in ecoinvent are really high, as pointed out in \cite{pirson2022environmental}.
However, for human toxicity, water-related ecotoxicity and freshwater eutrophication, the results obtained with the two databases vary by two to four orders of magnitude.

Inconsistencies of the results obtained with Sphera LCA and ecoinvent databases have been spotted by previous studies, for various type of case studies: packaging \cite{su12239948}, electric cars \cite{kalverkamp2020impacts}, agricultural products \cite{herrmann2015does},  buildings \cite{emami2019life}, etc.
According to \cite{su12239948}, "this is partly due to the fact that ecoinvent datasets often include more background processes than the corresponding GaBi [Sphera LCA] datasets", therefore leading to higher overall impacts with ecoinvent. In addition, toxicity and ecotoxicity metrics are known to provide more variations (several orders of magnitude) than global warming potential because they account for several thousand elementary flows, with different environmental mechanisms \cite{rosenbaum2018uncertainty} \cite{teehan2014integrative}.

In addition, we notice that the two databases have different assumptions in their modeling of industrial processes, especially regarding the emissions generated during the production of electronic components. For instance, we observe in our study that almost 100\% of the marine ecotoxicity reported with the ecoinvent database is due to heavy-metal emissions to water.
In Sphera LCA, however, marine ecotoxicity is dominated (77\%) by heavy-metal emissions to the air. In other words, the emissions to the sea modeled in Sphera LCA are so small that the main contributions to marine ecotoxicity are the emissions to the air reflected in the sea. In contrast, heavy metal emissions to the sea are assumed in the background models of ecoinvent.
This observation is consistent with \cite{su12239948} that claims that ecoinvent has a wider scope for background processes than Sphera LCA. Consequently, this may partly explain the four orders of magnitude between Sphera LCA and ecoinvent results.

Hence, mixing the two databases would likely not provide coherent results given the important discrepancies observed.
Nevertheless, even if there are inconsistencies between these databases, each one seems to be internally consistent. This was verified by comparing the impacts of the production of different components and devices, and by comparing electricity mixes estimated by each database. The same orders of differences were observed on the same impact categories. This is not surprising given that in both databases the processes are mostly modeled with the same background processes (electricity, raw materials, etc.), making them highly interdependent and favoring internal consistency in the databases.

To conclude, the present study highlights significant inconsistencies between the Sphera LCA and ecoinvent databases, and confirms the observations made by \cite{su12239948}, \cite{kalverkamp2020impacts} and \cite{emami2019life}.
Database inconsistencies mainly exist for toxicity-related indicators and water eutrophication but not for global warming potential which is the most studied impact category, especially in studies dealing with the environmental impacts of ICT.
Yet, more systematic attention should be paid to these indicators as they reflect risks to ecosystems and human health \cite{rosenbaum2018uncertainty}, even if they are usually not addressed in the literature on impacts assessment of cryptocurrency, and poorly commented in the broader literature on ICT impact assessment.
It may also help to explain (and maybe reduce) inconsistencies between databases.

\section{Limitations and uncertainties}\label{sec:limiations-and-uncertainty}

This section addresses the limitations and uncertainties of this LCA study.
It also discusses the scope of the conclusions that can be drawn from this study.

\subsection{Limitations and uncertainties in this LCA study}\label{subsec:uncertainties}

In spite of detailed foreground and background models, this LCA study has some limitations. 
First, the use of proprietary databases, motivated by the willingness to use high-quality background data, implies a lack of transparency and hence of reproducibility (even if this is still a very common issue  in the LCA community).

Then, as mentioned in Section~\ref{sec:methods}, some cut-off rules applies, causing a scope mismatch between the models of the Antminer S9 builds with the Sphera LCA and the ecoinvent databases. More precisely, we do not take into account the connectors in Sphera LCA, nor the crystal oscillators in ecoinvent.
However, the impacts of crystal oscillators account for less than 0.01\% of he total results for all categories when using Sphera LCA. Similarly, connectors account for less than 1.5 \% of all impacts when using ecoinvent. Therefore, the scope mismatch between the two implementations has only a marginal impact on our results and conclusions.
 
In addition, this LCA study is also subject to uncertainties.
First, we did not physically have access to all parts of the miner to build the foreground model of this study. This leads to uncertainties, especially regarding the size and the mass of some components (corresponding to epistemic and parameter uncertainties, as pointed out in \cite{hauschild2018life}, chapter 11).
In contrast, given that the BM1387B mining ASICs represent by far the majority of ICs in the entire miner (189 units), we got physical access to BM1378B ICs and measured their silicon die area using an accurate microscope in order to reduce the uncertainty of related parameters.
Second, we have no information about the geographic location of the production processes of the ASIC miner, so we assume globalized activities (corresponding to epistemic uncertainty). 
Third, there is no information on the actual industrial processes used during the manufacturing and assembly phases. This leads to some simplifications, especially in the case of custom models of component manufacturing (e.g., the fans in Sphera LCA and the casing) for which we have potentially missed some processes.
Fourth, the limited number of unit processes available in the databases (that are not always up-to-date) weakens the relevance of certain modeling choices, which is another source of epistemic uncertainty.
In case of the ecoinvent model, much of the data for 2021 was already present in the ecoinvent v2 database (i.e., before 2013) and was not updated individually, but rather through other linked processes. This can be verified for each unit process in the ecoinvent documentation.
Last, uncertainty is introduced when going from the foreground to the background model through the selection of unit processes and their subsequent scaling as there is no database process that perfectly models a given real component (corresponding to an uncertainty of choice).
All the choices and trade-offs made for choosing the \emph{closest} processes to reality are detailed in the supporting information SI 1 and SI 2.

To estimate the effect of parameter and choice uncertainties on our results, we carried out a sensitivity analysis using the Sphera LCA database with three different scenarios: low, typical and high scenarios.
They cover the uncertainty in the sizes and masses of components, in the scaling applied for modeling a given component from a unit process, and in the choice of unit processes in some cases. 
The results for these three scenarios are presented in Fig. \ref{fig:database-comparison}.
The difference between the upper and typical scenario, and between the typical and the lower scenario are always less than 20\% of overall result from the typical scenario.
Nevertheless, let us mention that the uncertainty associated with the foreground model and the choice of specific unit processes is small compared with the influence of the choice of database on the results.

\subsection{End-of-life consideration}\label{subsec:end-of-life}

This study does not consider the impacts due to the end-of-life of the miners. Therefore, we cannot conclude whether end-of-life is a significant phase for any impact category.
End-of-life is a potential issue for the environmental impacts of hardware, especially in the case of a rapidly renewing technology like crypto-mining hardware.
In addition, the ASICs in Bitcoin miners are designed for computing the SHA-256 algorithm for mining blocks of the Bitcoin blockchain. Therefore, such ASICs are so specific that they cannot be used for any other task \cite{de2021bitcoin}\cite{taylor2013bitcoin}, not even for mining a proof-of-work cryptocurrency with another algorithm in its protocol.

In this study, we do not quantify the end-of-life of an ASIC miners due to a lack of data. 
Indeed, data on e-waste is scarce and inaccessible in two different ways. 
Firstly, there is a lack of foreground data on e-waste collection and disposal. As an illustration, the Global e-waste monitor estimates that about 83\% of the global e-waste generated in 2019 was not documented \cite{forti2020global}.
For the remaining part that is documented and that goes through the formal disposal cycle, there is no global reporting about the recycling and disposal chains in each country, which makes it difficult to model the formal disposal of e-waste in countries with intensive mining activity.
And even less is known about the share of informal recycling and disposal.
Therefore, it is not easy to approximate the share of obsolete ASIC miners that goes through the different disposal scenarios (e.g., formal copper recycling, formal incineration, or informal landfill). 

On top of the lack of foreground data, background data is also lacking.
Indeed, if end-of-life scenarios are implemented with the current available processes in Sphera LCA and ecoinvent databases, many scenarios would hardly be consistent. In particular, \textit{electronics in landfill} is not modeled by the two databases whereas landfill would be the dominant scenario for countries with intensive mining activity like Kazakhstan \cite{index2023map}\cite{balde2021regional}.
The end-of-life of an ASIC miner may also cause important impacts on toxicity-related indicators, in scenarios involving informal landfill recycling \cite{singh2019toxicity}\cite{li2020environmental}. 
Hence, the lack of knowledge about the end-of-life is a major limitation for LCA in general, and for this work in particular.
We hope future studies may help to overcome the difficulties of assessing the end-of-life impacts of ICT.


\section{Discussion}
\label{sec:beyond-carbon}

This last section discusses the results of this study in a broader context.
First, we compare the multi-indicators results of this study about the production of an ASIC miner with different use-phase scenarios.
Then, we compare our results with the results of a previous study by reproducing them. Third, we adopt our LCA
study to a more recent miner in order to assess the potential variations in the production impacts between different ASIC miners.

\subsection{Comparison between production and use phases}\label{subsec:prod-vs-use}

\begin{figure*}
    \centering
    \includegraphics[width=0.6\textwidth]{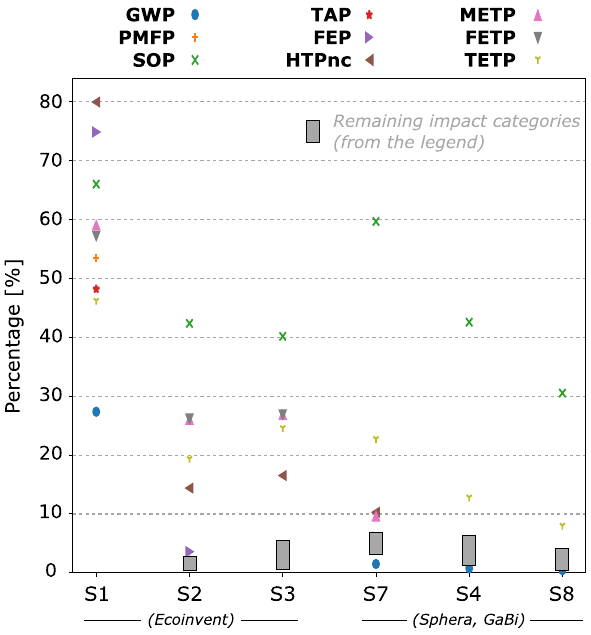}
    \caption{Share of impacts due to the production with respect to the sum of impacts from production and use phases, for the lifespan and electricity mix scenarios from Table \ref{table:use-phase-scenarios} and for different impact categories. We show the half of  impact categories with the highest share for the production. Data available in the Sheet ``data\_from\_figure\_5\_in\_manuscript'' of the Supporting Information SI 3.
    }
    \label{fig:production-vs-use}
\end{figure*}

For this analysis, we consider different use-phase scenarios for which we vary the electricity mix, the lifespan and the power consumption of the miner.
Then, we examine for which environmental indicators the production impacts are significant compared to the use phase impacts.
This approach allows us to compare the impacts from the production and the use phases, but not to draw conclusions on the whole life-cycle as the end-of-life is missing.

\begin{table*}
    \caption{Use-phase scenarios considered for comparison with the production phase}
    \label{table:use-phase-scenarios}
    \centering
    \begin{tabular}{|c|c|c|c|c|}
        \hline
          & Database  & Country   & \parbox{2.5cm}{\centering Power       \\consumption\\(kW)}  & \parbox{2.5cm}{\centering Lifespan\\(years)} \\
        \hline
        1 & Ecoinvent & Ca-Québec & 1.4                             & 3   \\
        2 & Ecoinvent & Global    & 1.4                             & 3   \\
        3 & Ecoinvent & USA       & 1.4                             & 3   \\
        4 & GaBi      & USA       & 1.4                             & 3   \\
        5 & GaBi      & USA       & 1.3                             & 3   \\
        6 & GaBi      & USA       & 1.5                             & 3   \\
        7 & GaBi      & USA       & 1.4                             & 1.5 \\
        8 & GaBi      & USA       & 1.4                             & 5   \\
        \hline
    \end{tabular}
\end{table*}

All the scenarios are summarized in Table \ref{table:use-phase-scenarios}. Regarding the electricity mix, we consider three cases:
\begin{itemize}
    \item the electricity mix of the USA, reported to have the highest mining activity to date, i.e., 37.4\% of the total in 2022 \cite{index2023map};
    \item a global electricity mix corresponding to a proportional weighting of electricity mix of countries where Bitcoin mining is reported to be significant in 2022 \cite{index2023map};
    \item the electricity mix of Québec, Canada, that illustrates a low-carbon and mainly renewable electricity mix.
          In fact, a significant Bitcoin mining activity takes place in Québec \cite{coinshare2022the}.

\end{itemize}

Regarding the power consumption, we consider three cases: 1.3 kW, 1.4 kW and 1.5kW. Vendors of the Antminer S9 indicates a power consumption of 1.372 kW. Higher estimates take into account some power losses, but we do not account for external active cooling systems and other general infrastructure.

Regarding the lifespan, we also consider three cases:
\begin{itemize}
    \item 3 years in the main scenario as the profitability duration of the Antminer S9 was estimated to be 3.39 years by \cite{de2021bitcoin}.
    \item 1.5 years should be a more realistic lifespan estimate of the majority of ASIC miners. Indeed, Antminer S9 is known as the Bitcoin miner with the longest lifespan, while \cite{de2021bitcoin} estimated an average lifespan of 1.29 years for ASIC miners on the period 2014-2021. 1.5 years is also the lifespan used in the baseline scenario of \cite{kohler2019life}.
    \item 5 years is an upper-bound of the miner lifespan. According to \cite{coinshare2022the}, Antminer S9 were still dominating the Bitcoin network in 2021 even if it was released in 2016.
          This is not incompatible with \cite{de2021bitcoin} as the sprike in Bitcoin price in 2021 has made some old and inefficient ASIC miners economically profitable again.
\end{itemize}

The results are shown in Fig. \ref{fig:production-vs-use}. Scenarios 5 and 6 are not shown because they were very similar to scenario 4.
Results with the energy mix of Quebec are really different from those of the other scenarios. Indeed, when using renewable energy for the cryptocurrency mining, the production phase of the miners becomes the hotspot of the activity for a large majority of impact categories.
The low-carbon and renewable energy mix scenario should be seriously taken into account: renewable energy represented 27\% of the Bitcoin network power consumption, and nuclear power 12 \% in December 2021 according to \cite{coinshare2022the}. Before China's ban on cryptocurrency mining in 2021, the share of renewable electricity seasonally exceeded 50\% of the consumption of the Bitcoin network \cite{de2022revisiting}.
Therefore, this shifts environmental costs from the computation phase of cryptocurrency mining to the production phase of mining equipment for almost all environmental impact categories.
Regarding all other scenarios, it should be noticed that production is responsible for 30\% to 70\% of mineral resource scarcity due to production plus use phases. This was expected because a large number of materials, with a high level of purity, are needed for electronics manufacturing \cite{williams20021}.

For all toxicity-related indicators (except for human cancerogenic toxicity), the production exceeds 15\% of the impacts in all scenarios with ecoinvent, and lies between 4 and 25\% in scenarios with Sphera LCA, when considering a 1.5 year lifespan.
Thus, production should be taken into account when discussing human and eco-toxicity of Bitcoin mining.

Finally, the production also ranges from 5 to 30\% of the impacts in all scenarios (except the Québec's one, in which it is higher) for ozone formation (human health), fine particulate matter formation, ozone formation (ecosystems), terrestrial ecotoxicity, terrestrial acidification and freshwater eutrophication.
This suggests that the contribution of production to these categories should also be analyzed when studying the impacts of Bitcoin mining.

\subsection{Comparison with another study}\label{subsec:kohler}

\begin{figure*}
    \centering
    \includegraphics[width=1\textwidth]{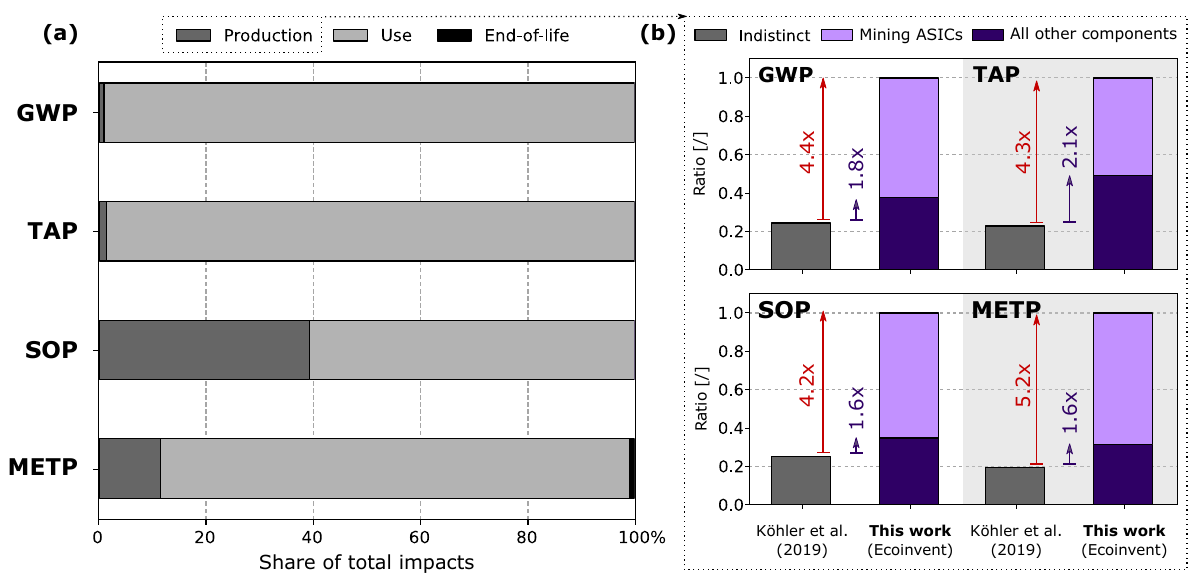}
    \caption{Comparison of our results with respect to the results from \cite{kohler2019life}: (a) \textit{Köhler et al.}'s results for the whole life-cycle, and (b) comparison with our results for the production phase only.
        The model of miner production from \textit{Köhler et al.} is depicted by an "indistinct" component.
        Only four impact categories are presented as the other categories behave in the same way. Data available in the Sheet ``data\_from\_figure\_6\_in\_manuscript'' of the Supporting Information SI 3.}
    \label{fig:kohler}
\end{figure*}

Thanks to its detailed supplementary material, we have been able to reproduce the results of \cite{kohler2019life} who also use the ecoinvent database and the ReCIPe method. It allows us (i) to compare our results with their study, (ii) to analyze the influence of considering specialized hardware on LCA results, and (iii) to compare the the life-cycle phases with a multi-indicators point of view, which is not the focus of \cite{kohler2019life}.
The results for four indicators are shown in Fig. \ref{fig:kohler}.

Figure \ref{fig:kohler}(b) compares LCA results obtained with ecoinvent in the present study with the results for the production of an ASIC miner in the model used by \cite{kohler2019life}. The model from \cite{kohler2019life} was generated with ecoinvent but, in contrast with our study, it does not take into account the specificity of mining hardware, and thus neglects the 189 ASICs present in the Bitmain Antminer S9 which largely contribute to the production impacts.
When excluding the ASICs, results of the two studies are in the same order of magnitude, from 1.4 to 2.2$\times$ higher for this work.
This suggests that if \cite{kohler2019life} take into account the specificity of an ASIC miner architecture, and especially the presence of numerous ASICs, their results should be revised upwards.
Thus, taking into account hardware specificity, even in a simplified way, may largely change the results for the production phase.

Figure \ref{fig:kohler}(a) shows the contribution of the different life-cycle phases of the Bitcoin estimated by \cite{kohler2019life} to the total impacts.
We recompute their results to compare production, use and end-of-life phases using the ReCiPe method (which is also used by the authors of \cite{kohler2019life}).
The end-of-life contribution seems to be marginal and the results are really similar to the ones presented in the Section \ref{subsec:prod-vs-use}.
As mentioned in the original paper, the production is negligible regarding the use phase when the focus is on the global warming.
However, this is not the case for other indicators like resource scarcity. These results are consistent with those of the present study.

\subsection{Extrapolation to another ASIC miner}\label{subsec:s19-pro}
This study presents an LCA of the Antminer S9 ASIC miner which is not representative of all ASIC miners. This choice is explained by its popularity and long lifespan.
However, this miner was released in 2016 and mining hardware has evolved since then.
In order to explore the evolution of hardware and the possible disparities in the embodied impacts of different ASIC miners, we adapted our LCA model to the Antminer S19 pro, a larger miner released in 2020.

With this simplified model, the total impacts of producing an Antminer S19 pro are estimated to be between 2 and 2.5$\times$ the impacts of Antminer S9 for all ReCiPe 2016 impact categories (details of the modeling changes are available in the support Information SI 1 and SI 2). This is a low-bound estimate as only some important parameters were scaled up.
It still indicates a clear difference of impacts between the two miners. This difference should be kept in mind when considering the environmental impacts of larger ASIC miners compared to the Antminer S9.
Thus, if larger miners becomes more common with the years, the constant energy efficiency improvements of cryptocurrency miners does not necessarily imply a reduction of the embodied impact of the mining equipment.
Hence, the impacts of production of one Antminer S9 evaluated in the present study are not representative for all ASIC miners machines and consequently cannot be directly scaled to the entire mining network.

\section{Conclusion}\label{sec:conclusion}
Cryptocurrencies are regularly pointed out for the massive energy consumption of their mining. However, crypto-mining also results in the production of millions of highly specialized and short-life ASIC miners per year.
In this article we presented the first LCA of the production of a specialized Bitcoin mining equipment. 
The ASICs emerges as the main contributor in almost all impact categories. The independent background modeling performed with Sphera LCA (GaBi) and ecoinvent databases highlighted important inconsistencies among these databases, especially on toxicity-related impact categories. Such a database mismatch, that was previously reported in other studies, calls for more attention and discussion on all impacts categories, beyond global warming potential.  
We also performed a comparison of our LCA results on the fabrication of an ASIC miner with various scenarios of use phase. We can conclude that, if the production impacts are negligible in comparison to the use phase ones regarding global warming, that is not the case for other indicators like mineral resources scarcity. We also noticed the important role played by the equipment lifetime and the mining location in such a hotspot analysis. Last but not least, the production impacts are really significant for almost all impact categories when mining is performed in a country with renewable and low-carbon electricity. This highlights the growing importance of production related impacts as the electricity used becomes less carbon-intensive.

Future work is needed to properly account for the end-of-life of the ASIC miners, and to interpret and reconcile the observed database discrepancy. Such future work goes well beyond the impact assessment of Bitcoin mining equipment.

\section*{Funding information}
This work was supported by ENS Rennes (France), Université catholique de Louvain (Belgium),  the Fonds européen de développement régional (FEDER) and the Wallonia within the Wallonie-2020.EU program.

\section*{Acknowledgments} We thank the ECS team members for their advice and reviewing, and especially Augustin Wattiez. We also thank Clément Morand for his support and reviewing.  

\section*{Conflict of Interest} The authors have no conflict of interest to disclose.
\section{Supporting Information}
Supporting Information SI 1 is the detailed inventory of the processes and quantities used in the LCA study associated with the components to be modeled.
The Supporting Information SI 2 is a detailed documentation of the foreground modeling supported by the analysis of pictures of the components.  
Supporting Information SI 3 gathers the data presented in figures.

\bibliography{biblography}

\end{document}